\def \invpb    {\relax\ifmmode{\rm pb^{-1}}\else{$\rm pb^{-1}$}\fi}
\def \u        {\relax\ifmmode{\mu}\else{$\mu$}\fi}
\def \epem     {\relax\ifmmode{e^+e^-}\else{$e^+e^-$}\fi}
\def \3l       {\relax\ifmmode{\ell^+\ell^-\ell}\else{$\ell^+\ell^-\ell$}\fi}
\def \lplm     {\relax\ifmmode{\ell^+\ell^-}\else{$\ell^+\ell^-$}\fi}
\def \lnu      {\relax\ifmmode{\ell\bar\nu_{\ell}}\else{$\ell\bar\nu_{\ell}$}\fi}
\def \upum     {\relax\ifmmode{\mu^+\mu^-}\else{$\mu^+\mu^-$}\fi}
\def \llb      {\relax\ifmmode{l\bar l}\else{$l\bar l$}\fi}
\def \emu      {\relax\ifmmode{e\mu}\else{$e\mu$}\fi}
\def \Upr      {\relax\ifmmode{\Upsilon'}\else{$\Upsilon'$}\fi}
\def \uones    {\relax\ifmmode{\Upsilon \rm (1S)}\else{$\Upsilon \rm (1S)$}\fi}
\def \utwos    {\relax\ifmmode{\Upsilon \rm (2S)}\else{$\Upsilon \rm (2S)$}\fi}
\def \uthrees  {\relax\ifmmode{\Upsilon \rm (3S)}\else{$\Upsilon \rm (3S)$}\fi}
\def \ufours   {\relax\ifmmode{\Upsilon \rm (4S)}\else{$\Upsilon \rm (4S)$}\fi}
\def \ufives   {\relax\ifmmode{\Upsilon \rm (5S)}\else{$\Upsilon \rm (5S)$}\fi}
\def \qqb      {\relax\ifmmode{q\bar q}\else{$q\bar q$}\fi}
\def \ttb      {\relax\ifmmode{t\bar t}\else{$t\bar t$}\fi}
\def \uub      {\relax\ifmmode{u\bar u}\else{$u\bar u$}\fi}
\def \ddb      {\relax\ifmmode{d\bar d}\else{$d\bar d$}\fi}
\def \ssb      {\relax\ifmmode{s\bar s}\else{$s\bar s$}\fi}
\def \ccb      {\relax\ifmmode{c\bar c}\else{$c\bar c$}\fi}
\def \bbb      {\relax\ifmmode{b\bar b}\else{$b\bar b$}\fi}
\def \ppb      {\relax\ifmmode{p\bar p}\else{$p\bar p$}\fi}
\def \BBB      {\relax\ifmmode{B\bar B}\else{$B\bar B$}\fi}
\def \pb       {\relax\ifmmode{\bar p}\else{$\bar p$}\fi}
\def \Wprime   {\relax\ifmmode{W^\prime}\else{$W^\prime$}\fi}
\def \Zprime   {\relax\ifmmode{Z^\prime}\else{$Z^\prime$}\fi}
\def \Wpm      {\relax\ifmmode{W^{\pm}}\else{$W^{\pm}$}\fi}
\def \Wp       {\relax\ifmmode{W^{+}}\else{$W^{+}$}\fi}
\def \Wm       {\relax\ifmmode{W^{-}}\else{$W^{-}$}\fi}
\def \Zz       {\relax\ifmmode{Z^0}\else{$Z^0$}\fi}
\def \photino  {\relax\ifmmode{\tilde{\gamma}}\else{$\tilde{\gamma}$}\fi}
\def \gluino   {\relax\ifmmode{\tilde{g}}\else{$\tilde{g}$}\fi}
\def \sbottom  {\relax\ifmmode{\tilde{b}}\else{$\tilde{b}$}\fi}
\def \stop     {\relax\ifmmode{\tilde{t}}\else{$\tilde{t}$}\fi}
\def \squark   {\relax\ifmmode{\tilde{q}}\else{$\tilde{q}$}\fi}
\def \slepton  {\relax\ifmmode{\tilde{\ell}}\else{$\tilde{\ell}$}\fi}
\def \snu      {\relax\ifmmode{\tilde{\nu}}\else{$\tilde{\nu}$}\fi}
\def \wino     {\relax\ifmmode{\tilde{W}}\else{$\tilde{W}$}\fi}
\def \zino     {\relax\ifmmode{\tilde{Z}}\else{$\tilde{Z}$}\fi}
\def \sfermion {\relax\ifmmode{\widetilde f}\else{$\widetilde f$}\fi}
\def \chichi   {\relax\ifmmode{\tilde{\chi}_1^\pm \tilde{\chi}_2^0}
                \else{${\tilde{\chi}_1^\pm \tilde{\chi}_2^0}$}\fi}
\def \chizero  {\relax\ifmmode{\tilde{\chi}_1^0}\else{$\tilde{\chi}_1^0$}\fi}
\def \chipm	{\relax\ifmmode{\tilde{\chi}^{\pm}}
                \else{$\tilde{\chi}^{\pm}$}\fi}
\def \chizero	{\relax\ifmmode{\tilde{\chi}^0}\else{$\tilde{\chi}^0$}\fi}
\def \chione   {\relax\ifmmode{\tilde{\chi}_1^\pm}
                \else{$\tilde{\chi}_1^\pm$}\fi}
\def \chionep  {\relax\ifmmode{\tilde{\chi}_1^+}\else{$\tilde{\chi}_1^+$}\fi}
\def \chionem  {\relax\ifmmode{\tilde{\chi}_1^-}\else{$\tilde{\chi}_1^-$}\fi}
\def \chitwo   {\relax\ifmmode{\tilde{\chi}_2^0}\else{$\tilde{\chi}_2^0$}\fi}
\def \lpm      {\relax\ifmmode{\ell^\pm}\else{$\ell^\pm$}\fi}
\def \lplm     {\relax\ifmmode{\ell^{+}\ell^{-}}\else{$\ell^{+}\ell^{-}$}\fi}
\def \llbar    {\relax\ifmmode{\ell \bar{\ell}}\else{$\ell \bar{\ell}$}\fi}
\def \lnu      {\relax\ifmmode{\ell^{\pm}\nu}\else{$\ell^{\pm}\nu$}\fi}
\def \wlnu     {\relax\ifmmode{W^{\pm}\rightarrow\ell^{\pm}\nu}
                \else{$W^{\pm}\rightarrow\ell^{\pm}\nu$}\fi}
\def \wenu     {\relax\ifmmode{W^{\pm}\rightarrow e^{\pm}\nu}
                \else{$W^{\pm}\rightarrow e^{\pm}\nu$}\fi}
\def \wmunu    {\relax\ifmmode{W^{\pm}\rightarrow\mu^{\pm}\nu}
                \else{$W^{\pm}\rightarrow\mu^{\pm}\nu$}\fi}
\def \zll      {\relax\ifmmode{Z^0\rightarrow \ell^+\ell^-}
                \else{$Z^0\rightarrow \ell^+\ell^-$}\fi}
\def \zee      {\relax\ifmmode{Z^0\rightarrow e^+e^-}
                \else{$Z^0\rightarrow e^+e^-$}\fi}
\def \zmumu    {\relax\ifmmode{Z^0\rightarrow\mu^+\mu^-}
                \else{$Z^0\rightarrow\mu^+\mu^-$}\fi}
\def \ztautau  {\relax\ifmmode{Z^0\rightarrow\tau^+\tau^-}
                \else{$Z^0\rightarrow\tau^+\tau^-$}\fi}
\def \tautau   {\relax\ifmmode{\tau^+\tau^-}\else{$\tau^+\tau^-$}\fi}
\def \mumu     {\relax\ifmmode{\mu^+\mu^-}\else{$\mu^+\mu^-$}\fi}
\def \ee       {\relax\ifmmode{e^+e^-}\else{$e^+e^-$}\fi}
\def \eee      {\relax\ifmmode{e^+e^-e^+}\else{$e^+e^-e^+$}\fi}
\def \eemu     {\relax\ifmmode{e^+e^-\mu^+}\else{$e^+e^-\mu^+$}\fi}
\def \emumu    {\relax\ifmmode{e^+\mu^-\mu^+}\else{$e^+\mu^-\mu^+$}\fi}
\def \mumumu   {\relax\ifmmode{\mu^+\mu^-\mu^+}\else{$\mu^+\mu^-\mu^+$}\fi}
\def \qbar     {\relax\ifmmode{\bar{q}}\else{$\bar{q}$}\fi}
\def \qsq      {\relax\ifmmode{\rm q^2}\else{$\rm q^2$}\fi}

\def \etl      {\relax\ifmmode{\em et al.,   \rm}\else{$\em et al.,   \rm$}\fi}
\def \psp      {\relax\ifmmode{\,}\else{$\,$}\fi}
\def \PT       {\relax\ifmmode{P_{T}}
                \else{$P_{T}$}\fi}
\def \ET       {\relax\ifmmode{E_{T}}
                \else{$E_{T}$}\fi}
\def \MET      {\relax\ifmmode{\mbox{$\raisebox{.3ex}{$\not$}\ET$}}
                \else{$\mbox{$\raisebox{.3ex}{$\not$}\ET$}$}\fi}
\def \ete      {\relax\ifmmode{E_{T}^{e}}\else{$E_{T}^{e}$}\fi}
\def \ptm      {\relax\ifmmode{P_{T}^{\mu}}\else{$P_{T}^{\mu}$}\fi}
\def \mgev     {GeV/$c^{2}$}
\def \pgev     {GeV/$c$}
\def \sp       {\relax\ifmmode{\;}\else{$\;$}\fi}
\def \dk       {\relax\ifmmode{\rightarrow}\else{$\rightarrow$}\fi}
\def \chisq    {\relax\ifmmode{\chi^2}\else{$\chi^2$}\fi}
\def \chisqs#1 {\relax\ifmmode{\rm \chi^2_{#1}}\else{$\rm \chi^2_{#1}$}\fi}
\newcommand    {\Romnum}[1]{\uppercase\expandafter{\romannumeral #1}}
\def \eff      {\relax\ifmmode{\varepsilon}\else{$\varepsilon$}\fi}
\def \stimes   {\relax\ifmmode{\vec{\times}}\else{$\vec{\times}$}\fi}

\def \gtsim    {\relax\ifmmode{\mathrel{\mathpalette\oversim >}}
                  \else{$\mathrel{\mathpalette\oversim >}$}\fi}
\def \ltsim    {\relax\ifmmode{\mathrel{\mathpalette\oversim <}}
                  \else{$\mathrel{\mathpalette\oversim <}$}\fi}
\def\oversim#1#2{\lower4pt\vbox{\baselineskip0pt \lineskip1.5pt
            \ialign{$\mathsurround=0pt#1\hfil##\hfil$\crcr#2\crcr\sim\crcr}}}

\newcommand{\ba}{\begin{array}}
\newcommand{\ea}{\end{array}}
\newcommand{\bc}{\begin{center}}
\newcommand{\ec}{\end{center}}
\newcommand{\bn}{\begin{enumerate}}
\newcommand{\en}{\end{enumerate}}
\newcommand{\bq}{\begin{equation}}
\newcommand{\eq}{\end{equation}}
\newcommand{\bi}{\begin{itemize}}
\newcommand{\ei}{\end{itemize}}
\newcommand{\bh}{\begin{math}}
\newcommand{\eh}{\end{math}}
\newcommand{\br}{\begin{flushright}}
\newcommand{\er}{\end{flushright}}
\newcommand{\bl}{\begin{flushleft}}
\newcommand{\el}{\end{flushleft}}
\newcommand{\bt}{\begin{tabular}}
\newcommand{\et}{\end{tabular}}

\newcommand{\bHuge}{\begin{Huge}}
\newcommand{\bhuge}{\begin{huge}}
\newcommand{\bLARGE}{\begin{LARGE}}
\newcommand{\bLarge}{\begin{Large}}
\newcommand{\blarge}{\begin{large}}
\newcommand{\eHuge}{\end{Huge}}
\newcommand{\ehuge}{\end{huge}}
\newcommand{\eLARGE}{\end{LARGE}}
\newcommand{\eLarge}{\end{Large}}
\newcommand{\elarge}{\end{large}}

\documentstyle[11pt,epsf]{article}

\begin{document}
\baselineskip 11pt
$\;$
\vspace{4cm}
\begin{center}
\LARGE{\bf NEW PARTICLE SEARCHES AT TEVATRON (II)}
\end{center}

\begin{center}
\rm{TERUKI KAMON} \\
\rm{Department of Physics, Texas A\&M University}\\
\rm{College Station, Texas 77843-4242}\\[.2in]
\rm{On behalf of the CDF and D\O\sp Collaborations}\\[8cm]
\end{center}
 
\begin{abstract}
Various recent results of new particle searches at the Fermilab Tevatron
are presented. No evidence is found for 
supersymmetric particles (chargino, gluino),
leptoquark bosons and
heavy gauge bosons in $\ppb$ collisions at $\sqrt{s}$ = 1.8 TeV.
Excluded mass regions for each particle are determined.
\end{abstract}

\newpage
\baselineskip 14pt
\section{Introduction}

With the data from
CERN's LEP electron-positron collider, DESY's HERA electron-proton collider and
Fermilab's Tevatron proton-antiproton collider,
the Standard Model (SM) has received overwhelming
experimental support.
No new physics has been observed essentially at mass below $\frac{1}{2} M_Z$.
Thus, it is imperative that the Fermilab Tevatron,
the world's highest energy particle collider,
be used for searches of new phenomena.
The CDF and D\O\sp experiments have been actively studying their data
for evidence of previously unobserved particles.
Presented below are the latest results of searches at the Tevatron
for supersymmetric particles (chargino, gluino),
leptoquark bosons and heavy gauge bosons.

\section{Search for $\chipm_1 \chizero_2$ 
	and $\gluino \gluino$ at CDF and D\O\ }

The Minimal Supersymmetric Standard Model (MSSM) \cite{MSSM}
is a {\it supersymmetrized} SM with two Higgs doublets,
which is one of the most appealing theories proposed to test
grand unification \cite{SUSY-GUT}.
Conservation of $R$-parity requires the SUSY particles to be produced in pairs
and prevents decays of the lightest supersymmetric particle 
(lightest neutralino $\chizero_1$).
The most distinctive signatures of 
chargino-neutralino ($\chipm_1 \chizero_2$) and
$\gluino \gluino$ ($\squark\squark$, $\gluino \squark$) 
pair production are 
trilepton events \cite{Trilepton} and
like-sign (LS) dileptons 
associated with large missing transverse energy ($\MET)$ plus multi-jets
\cite{LS-Dilepton}.

Pair-produced $\chipm_1 \chizero_2$ 
decay as $\chione\rightarrow \lpm\nu\chizero$ and
$\chitwo\rightarrow \lplm \chizero$.
The striking signature of these events is thus three isolated
leptons plus $\MET$.
The CDF search is based on 100 \invpb data.
Most of the events are from
inclusive $e$ and $\mu$ triggers at $p_T \sim$ 10 \pgev.
The trilepton requirements are
$p_T(\ell_1) >$ 11 \pgev\sp and $p_T(\ell_{2,3}) >$ 5 (4) \pgev\sp for
$e$ ($\mu$), which is the same as
in the previous analysis (based on 19 \invpb) \cite{CDF_TL_1A}.
Eight events are left which are consistent with the SM background.
A further cut of $\MET >$ 15 GeV is imposed to reduce the background.
There is no trilepton event candidate found,
which is consistent with an expected background of $0.4\pm0.1$
for four trilepton ($eee$, $ee\mu$, $e\mu\mu$, $\mu\mu\mu$) modes.
The D\O\ trilepton analysis requires a lower cut of
$p_T(\ell_{1,2,3}) >$ 5 \pgev,
although the events are from several trigger paths
(inclusive $e$ and $\mu$ samples and $ee$, $e\mu$ and $\mu \mu$ dilepton 
samples).
The acceptance difference due to those trigger thresholds are corrected
using a Monte Carlo simulation.
D\O\sp also observes no trilepton event candidate in 14 \invpb, which
complies with an expected background of 
$0.8\pm0.5$ for $eee$, $0.8\pm0.4$ for $ee\mu$, 
$0.6\pm0.2$ for $e\mu\mu$, $0.1\pm0.1$ for $\mu\mu\mu$ \cite{D0_TL_1A}.

Both CDF and D\O\sp have a similar constraint inspired by
supergravity models \cite{SUGRA}.
The CDF analysis \cite{CDF_TL_1A} 
calculates slepton ($\slepton$) and sneutrino ($\snu$) masses 
	from $\tan\beta$, $M(\gluino)$, and 
	$M(\squark)\,$ using the renormalization group equations~\cite{RGE},
while D\O\sp \cite{D0_TL_1A} uses a minimal supergravity model
in ISAJET \cite{ISAJET}.
In those models,
chargino and neutralino have three-body decays.
The CDF analysis excludes
$M(\chipm_1) <$ 66 \mgev\sp (95\% C.L.)
at $\tan\beta$ = 2, $M(\squark) = 1.05 M(\gluino)$ and $\mu = -400$ GeV
(the region of maximum experimental sensitivity).
The CDF limit on $\chipm_1$ (66 \mgev) 
is comparable to the LEP1.5 result~\cite{LEP150}.
The D\O\ result is not sensitive to set any mass limits for the ISAJET model.
Limits on $\sigma \cdot BR$(4 trilepton modes) are also obtained:
0.6 pb (CDF) and 4 pb (D\O) for 70-\mgev\sp chargino.

\begin{figure}[t]
\epsfysize=3.5in
\center{\leavevmode\epsffile{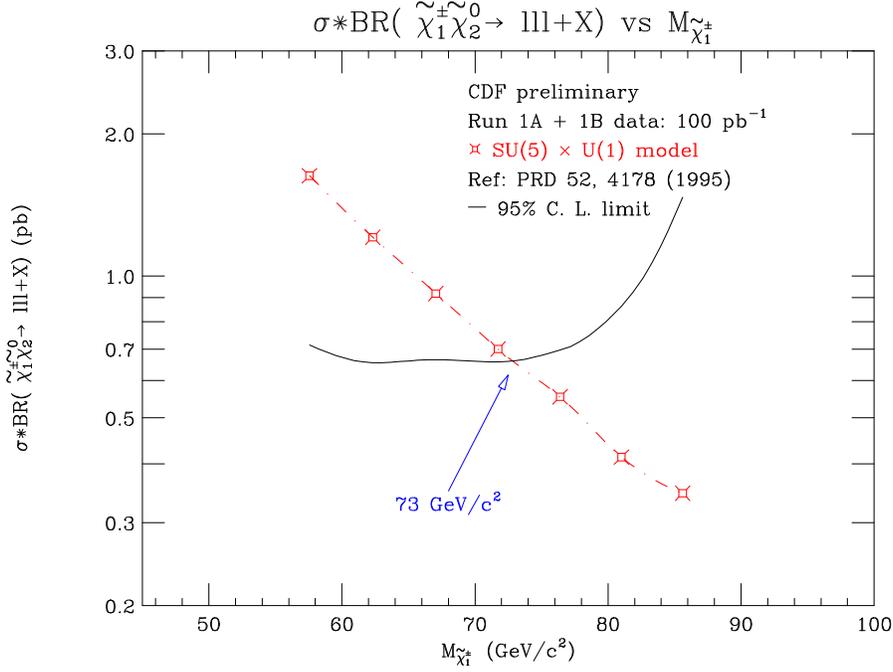}}
\caption{95\% C.L. upper limit on
$\sigma \cdot BR(\chichi\rightarrow 3 \ell X)$ vs. $M(\chione$).  The points
are the predictions of ISAJET. 
Note that 
$BR(\chichi\rightarrow 3 \ell X)$ = 
$BR(\chipm_1 \rightarrow \ell^{\pm} \nu \chizero_1) \cdot
 BR(\chizero_2 \rightarrow \slepton^{\pm}_R \ell^{\mp} ) \cdot
 BR(\slepton_R^{\pm} \rightarrow \ell_{R}^{\pm} \chizero_1)$ 
	for four trilepton modes.}
\label{upp_lim}
\end{figure}

CDF also examines one particular supergravity model, flipped SU(5) model
\cite{SU5xU1}.
In this model, $M(\slepton_R) < M(\chizero_2) < M(\slepton_L)$,
so that the trilepton signal is nearly maximized via
$BR(\chizero_2 \rightarrow \slepton^{\pm}_R \ell^{\mp}) \sim$ 66\% 
($e$ and $\mu$) and
$BR(\slepton_R^{\pm} \rightarrow \ell_{R}^{\pm} \chizero_1)$ = 100\%.
However, the mass difference of $\slepton_R$ and $\chizero_1$ decreases
as $M(\chipm_1)$ increases, so that
the total trilepton acceptance as a function of $M(\chipm_1)$ becomes
flat at about 5\% at 60 \mgev\sp and falls off
for $M(\chipm_1) \gtsim$ 75 \mgev.
Figure~\ref{upp_lim} shows the 95\% C.L. upper limit curve on
$\sigma \cdot BR(\chichi\rightarrow 3 \ell X)$.
The points in the figure are the ISAJET predictions
for $\chichi$ production in the supergravity model.
The limit on $M(\chipm_1)$ is 73 \mgev.

The LS dilepton approach is complementary to the classic
multi-jets+$\MET$ analysis in the search for $\gluino\gluino$,
$\gluino \squark$ and $\squark \squark$ production.
The signature arises mainly from $\gluino \gluino$ production
followed by dominant decays of 
$\gluino \rightarrow q \bar{q^{\prime}} \chipm_1$ and
$\gluino \rightarrow q \bar{q} \chizero_2$
if $M(\squark) > M(\gluino)$.
Thus, the final state contains two or more leptons
(from $\chipm_1$ and $\chizero_2$ decays), $\MET$ and jets.
Since the gluino is a Majorana fermion, there is no charge correlation
between these leptons.
Such a data analysis begins with a dilepton sample
($p_{T}(\ell_{1(2)}) >$ 12 (10) \pgev)
and subsequently requires $\MET>$ 25~GeV and $N(j) \geq 2$ with
$E_T(j_{1(2)}) >$ 15 GeV in $| \eta (j_{1(2)}) | < 1.1 (2.4).$
Since the production cross section times branching ratio is small,
CDF searched for the dilepton signature without the LS requirement.
In 19 \invpb, one event ($\mu^+\mu^-$) is observed, which
is consistent with an expected background of
$2.39^{+0.99}_{-0.76}$.
The lower gluino mass at
95\% C.L. is calculated to be 154 \mgev\sp for $M(\squark) \gg M(\gluino)$
and 224 \mgev\sp for $M(\squark) = M(\gluino)$
at $\tan \beta = 4$ and $\mu = -400$ GeV \cite{CDF_LS_1A}.
Those limits weakly depend on the $\tan\beta$ values (2-8) and
the $\mu$ values (200 GeV $< |\mu| <$ 1000 GeV).

\section{Search for Leptoquark Bosons at CDF and D\O }

Leptoquark is a generic term for color-triplet bosons which couple both
to a quark and a lepton.
They appear in many SM extensions which join the quark and lepton sectors
at more fundamental levels \cite{LQ_Models}.
Both CDF and D\O\sp search for pair-produced lepton quarks.
The signatures are $\ell^{+}\ell^{-}$ + 2 jets and $\ell$ + $\MET$ + 2 jets.

The D\O\sp analyses of the first and second generation
scalar leptoquarks ($LQ1$ and $LQ2$) searches
using $\lplm jj$ and $\ell \nu jj$ events are based on
15 \invpb data \cite{D0_LQ1,D0_LQ2}.
Neither analysis finds evidence for a leptoquark signal.
The mass limits are 130 (116) \mgev\sp at $\beta = 1~(0.5)$ for $LQ1$
with HMRS-B;
119 (97) \mgev\sp at $\beta = 1~(0.5)$ for $LQ2$ with CTEQ2pM.
Here $\beta$ = $BR(LQ_{i} \rightarrow \ell_{i} q_{i})$.
Note that MT-LO is the nominal choice in D\O's $LQ1$ and $LQ2$ analyses.
The CDF results on the $LQ1$ and $LQ2$ searches are based on
4.1 \invpb \cite{CDF_LQ1_1A} and 67 \invpb data \cite{CDF_NP_PbarP}.
The limits at $\beta = 1~(0.5)$ are 
113 (80) \mgev\sp for $LQ1$ with HMRS-B and
180 (141) \mgev\sp for $LQ2$ with CTEQ2pM.

CDF also searches for the third generation leptoquark ($LQ3$) in
$\tau^+ \tau^- j j$ mode, where one of the $\tau$ leptons
decays semileptonically
and the second $\tau$ has 1 or 3-prong hadronic decays.
The $\tau$-decay lepton ($e$ or $\mu$) is required to have
$p_T >$ 20 \pgev\sp ($|\eta| < 1.1$).
The hadronic decay $\tau$ must have uncorrected $E_T >$ 15 GeV
($|\eta| < 1$) associated 1 or 3 tracks
(the leading track $p_T >$ 10 \pgev)
in a $10^{\circ}$ cone around the energy cluster.
The charge of the $\tau$ candidate is defined to be the total charge
of 1 or 3 tracks and required to be an opposite sign to the lepton charge.
This reduces the QCD background by a factor of $\sim$2.
The azimuthal separation between the lepton and $\MET$ directions
should be less than $50^{\circ}$ to reduce the $W(\rightarrow \ell \nu)$
+ jets events.
Finally, two jets are required with
uncorrected $E_T >$ 10 GeV in $|\eta| < 4.2$.
No events are left; consistent with the expected SM background of
$1.2\pm^{1.0}_{0.2}$ events including 1.0 $Z \rightarrow \tau \tau$ event.

\begin{figure}[t]
\vspace{-1.5in}
\epsfysize=7.0in
\center{\leavevmode\epsffile{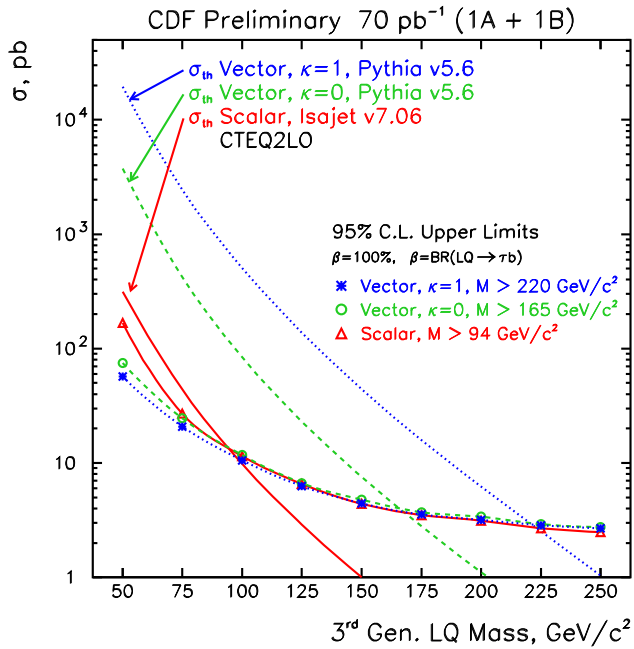}}
\vspace{-2.in}
\caption{95\% C.L. upper limit on
 $\sigma$ vs. $M(LQ3$) for $\beta$ = $BR(LQ3 \rightarrow \tau b)$ = 1.}
\label{Fig:CDF_LQ3}
\end{figure}

Figure~\ref{Fig:CDF_LQ3} shows the results of the mass limits on $LQ3$
for the scalar boson and the vector bosons (both $\kappa$ = 0 and 1, where
$\kappa$ is an anomolous chromomagnetic moment). 
All limits on $M(LQ3)$ are obtained at $\beta$ = 1.

\section{Search for $\Wprime \rightarrow WZ \rightarrow e \nu jj$ at CDF}

In extended gauge models \cite{EGM} proposed to
restore left-right symmetry to the weak force,
the right-handed, charged bosons $\Wprime$ can decay with large probability to
right-handed  $\ell_R \bar{\nu}_R$ pairs.
Both D\O\sp and CDF have searched for 
the heavy $\Wprime$ through the process $\ppb \rightarrow \Wprime \rightarrow e \nu$
(or $e_R \nu_R$ if $M(\nu_R) \ll M(\Wprime))$ in
74.4 \invpb and 19.7 \invpb, respectively.
The searches were made by assuming
a standard strength of the coupling and
the decay $\Wprime \rightarrow W Z$ to be 
suppressed by a left-right mixing angle
$\xi$ = $C$ [$M(W)/M(\Wprime)$]$^{2}$ where $C$ is ${\cal O}$(1).
The limits are
720 \mgev\sp \cite{D0_Wprime_enu} and 652 \mgev\sp \cite{CDF_Wprime_enu}.

The CDF experiment is conducting a complementary search for
$\Wprime$ in the decay to $W Z \rightarrow e \nu j j$
with 
$E_T(e) >$ 30 \mgev\sp ($| \eta | <$ 1.05), $\MET >$ 30 GeV,
$E_T(j_{1(2)}) >$ 50 (20) GeV.
Additional tracking, isolation and electron identification
criteria are also imposed.
The previous result \cite{CDF_NP_PbarP} is updated using 110 \invpb data:
512 $\Wprime$ candidate events are left.
We observe 7 events for $M(W+jj) >$ 600 \mgev, compared to
4.2 expected events.
We find no significant evidence for excess $Z$ production
produced in association with a $W$.

\begin{figure}[t]
\epsfysize=4.5in
\center{\leavevmode\epsffile{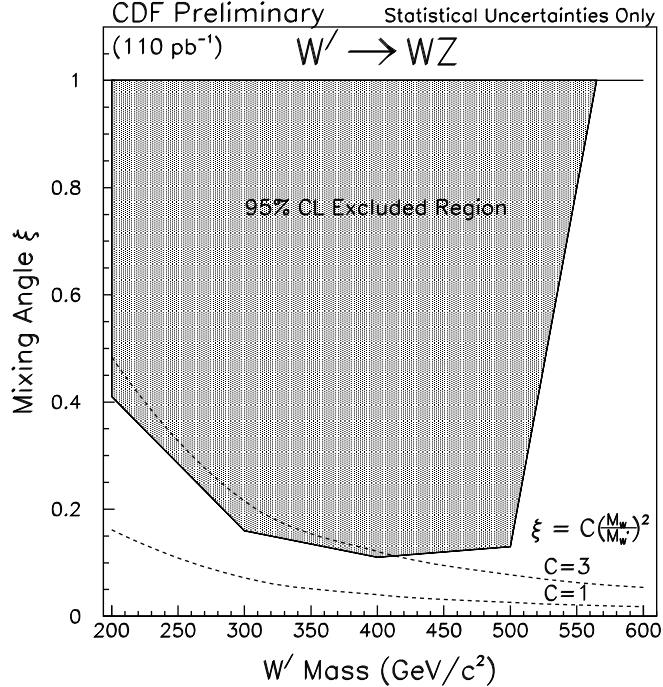}}
\vspace{-0.5in}
\caption{95\% C.L. upper limit excluded region of
	$\sigma \cdot BR(\Wprime \rightarrow WZ) \cdot BR(W \rightarrow e \nu)$
	for $\xi$ vs. $M(\Wprime)$.
	Note that systematic uncertainties are not included.
	The dashed lines with $C$ = 1 and 3 are for illustration purposes
	only to show what an excluded region vs. $C$ would look like. }
\label{Fig:CDF_Wprime_WZ}
\end{figure}

The limits are obtained by a likelihood fit with the data, background and
signal shapes.
The results of the excluded region in the $\xi$-$M(\Wprime)$ plane
are shown in Fig.~\ref{Fig:CDF_Wprime_WZ}.
The range of 200-560 \mgev\sp is excluded at $\xi$ = 1,
which is consistent with the limits above.
However, the CDF analysis is insensitive to setting any limit at $C$ = 1.

\section{Search for $\Zprime \rightarrow \lplm$ at CDF and D\O\ }

Neutral gauge bosons in addition to the $Z^{0}$ are expected in many
extensions of the SM \cite{Zprime_Model}.
These models typically specify the strength of the couplings
of such bosons to quarks and leptons but make no mass prediction.
$\Zprime$ bosons may be observed directly via
their decay to lepton pairs ($e^{+}e^{-}$, $\mu^{+} \mu^{-}$).
The CDF and D\O\sp results presented below
are derived assuming SM coupling strengths.

For $\Zprime \rightarrow ee$,
the CDF criteria are $E_T(e) >$ 25 GeV in $| \eta (e_{1(2)}) | < 1.1 (2.4)$.
The D\O\ requirements are similar:
$E_T(e) >$ 30 GeV in $| \eta (e_{1(2)}) | < 1.1 (2.5)$.
Additonal tracking, electron identification and isolation cuts are imposed.
The only appreciable background is from Drell-Yan $\gamma$ in the high mass
region.
For $M(ee) >$ 250 \mgev,
CDF (D\O) observes 8 (1) events in 67.6 (14.4) \invpb data
which is consistent with the expected background of 8.1 (1.4) events.

For $\Zprime \rightarrow \mu \mu$,
the CDF criteria are $p_T(\mu) >$ 20 GeV in $| \eta (\mu_{1(2)}) | < 0.6 (1.1)$
with additional quality cuts.
In 67 \invpb data,
seven events with $M(\mu\mu) >$ 200 \mgev\sp are observed consistent
with the expected rate from Drell-Yan production.

The mass limits are extracted by a binned maximum likelihood analysis on
the data comparing the data to a sum of the Drell-Yan background
and the $\Zprime$ expectation.
The Drell-Yan and $\Zprime$ distributions are modeled by the leading order
Monte Carlo. 
The CDF mass limits on $\Zprime$ are 620 \mgev\sp for $e^+e-$ and
590 \mgev\sp for $\mu^+ \mu^-$.
At this time, a combined limit of 650 \mgev\sp is available
from 67.6 \invpb $e^+e^-$ data and 
	71.3 \invpb $\mu^+ \mu^-$ data \cite{CDF_NP_PbarP}.
The D\O\sp mass limit from $e^+ e^-$ data is 490 \mgev\sp \cite{D0_Zprime_ee}.

\section{Summary}

The CDF and D\O\sp experiments have searched for various new physics
phenomena.
In these studies, there is no evidence for physics beyond the Standard
Model.

%
 
\begin{small}

\end{small}

\end{document}